# Bayesian optimization based on element mapping to design high-capacity NASICON-type cathode in sodium-ion battery

Sanghyeon Park[1, 2], Yoonsu Shim[1], Junpyo Hur[1], Sanghyeon Ji[1], Dongmin Jeon[2], Jong Min Yuk[1*] and Chan-Woo Lee[2*]

1 Department of Materials Science and Engineering, Korea Advanced Institute of Science and Technology, Daejeon 34141, Republic of Korea

2 Energy AI & Computational Science Laboratory, Korea Institute of Energy Research, Daejeon 34129, Republic of Korea

**Abstract**

To discover novel materials with high performance, there have been many attempts to adopt Bayesian optimization (BO) to materials science, owing to its efficiency in navigating complex and high-dimensional design spaces. However, the application of BO to material design has been suffered from handling discrete input variables, such as elements. Here, we introduce a novel element mapping strategy that encodes elemental identities into chemically meaningful continuous values, enabling to create easy-to-predict chemical spaces. We apply this new framework to design high capacity $Na_3V_2(PO_4)_2F_3$ (NVPF) cathode materials for sodium-ion batteries, targeting that shift all working voltages into the desired operational voltage window. The proposed framework successfully suggests 16 optimal element composition within 50 iterations. Our results demonstrate the way to overcome the limitation of categorical input that will likely broaden the applicability of BO to a wider range of material discoveries.

## Introduction

In response to the growing demand to discover new high-performance materials, traditional trial-and-error-based grid search approaches with experiments face significant limitations due to their high cost and extensive time requirement. To reduce the time and cost associated with these experiments, computational approaches such as density functional theory (DFT) calculations have been introduced[1, 2]. In particular, high-throughput screening with DFT calculations has emerged as a method to rapidly explore large search spaces and identify promising candidates materials[3]. Recently, remarkable advances in machine learning (ML), capable of direct prediction of materials properties from their structural information, have led to a new era of data-driven materials design[4-7].

Depending on how features are extracted from structural data, many ML techniques have been developed. Among them, deep learning models based on graph neural networks (GNNs) have shown high accuracy. The GNN models represent atomic structures as graphs composed of node and edges, where nodes correspond to individual atoms and edges correspond to chemical bonds. By explicitly learning these atomic relationships from structural data, the GNN models effectively capture complex interactions between atoms, enabling accurate predictions of key materials properties such as formation energy, binding energy, and electronic bandgap. Representative models including CGCNN[6] and M3GNet[7] achieve state-of-the-art predictive performance by training on large datasets, thus inherently relying on pre-existing datasets and requiring substantial computation for initial data acquisition.

Despite advances in property prediction, many studies still rely on grid search, which inevitably involves exhaustive evaluation of the search space, and thus carries fundamental inefficiencies. As design variables increase, the search space grows exponentially, making this approach increasingly prohibitive—especially since many sampled points may be irrelevant to the target.

In contrast to grid search, Bayesian optimization (BO) offers significantly improved efficiency in discovering optima by intelligently proposing evaluation points, thereby reducing the number of required evaluations in high-dimensional search spaces[8-10]. BO utilizes a probabilistic machine learning model—often Gaussian processes (GP)[11]—as a surrogate to predict the objective function and estimate the uncertainty with the observations. Based on the results of the surrogate model, the acquisition function proposes the next evaluation point.

Such advantages make BO an attractive method for materials research that requires efficient optimization over complex and high-dimensional search spaces. For instance, BO has been successfully applied to optimize high-entropy alloy (HEA) compositions for electrocatalytic oxygen reduction reactions[12] and multi-objective optimization of electrode manufacturing parameters in lithium-ion batteries[13].

However, despite its efficiency, the direct application of BO to designing new materials—such as discovering novel combinations for doping elements—has remained largely unexplored. This limitation arises primarily because Gaussian processes (GP), one of the most commonly employed surrogate models in BO, inherently assume continuous input variables, making it challenging to handle categorical or discontinuous variables such as elements. Although alternative surrogate models exist, they also struggle to handle high-dimensional categorical variables[14, 15].

To overcome this critical barrier, we introduce a novel element mapping strategy that uniquely encodes elemental identities into chemically meaningful continuous values. Unlike conventional categorical encoding approaches such as one-hot encoding, ordinal encoding, or embedding-based encodings, our element mapping explicitly preserves the intrinsic chemical characteristics of elements. This enables BO to systematically and efficiently explore previously inaccessible chemical space (search space) for materials discovery.

We apply this advanced material design strategy to design cathode materials for sodium-ion batteries (SIBs), an emerging alternative to lithium-ion batteries due to their low cost and abundant sodium resources[16-18]. Among sodium superionic conductor (NASICON) structures[19-21], $Na_3V_2(PO_4)_2F_3$ (NVPF) has attracted considerable attention for its high-rate capability and high working voltage[22, 23]. However, the low capacity of NVPF remains a critical challenge. Here, we target the discovery of novel element combinations that replace vanadium in NVPF ($Na_3M1_{(2.00-y)}M2_y(PO_4)_2F_3$, NMPF), enabling thermodynamically stable sodium-excess phases with the high capacity. By assigning chemically intuitive continuous values to elements, we construct a predictable and well-structured chemical space, significantly facilitating the efficient discovery of optimal element combinations with minimal evaluations. Our approach not only provides critical insights into designing high-energy-density cathode materials but also highlights the broad applicability of our BO-guided framework for efficiently discovering optimal element combinations, well beyond the limitations of conventional grid search or deep learning-based screening methods.

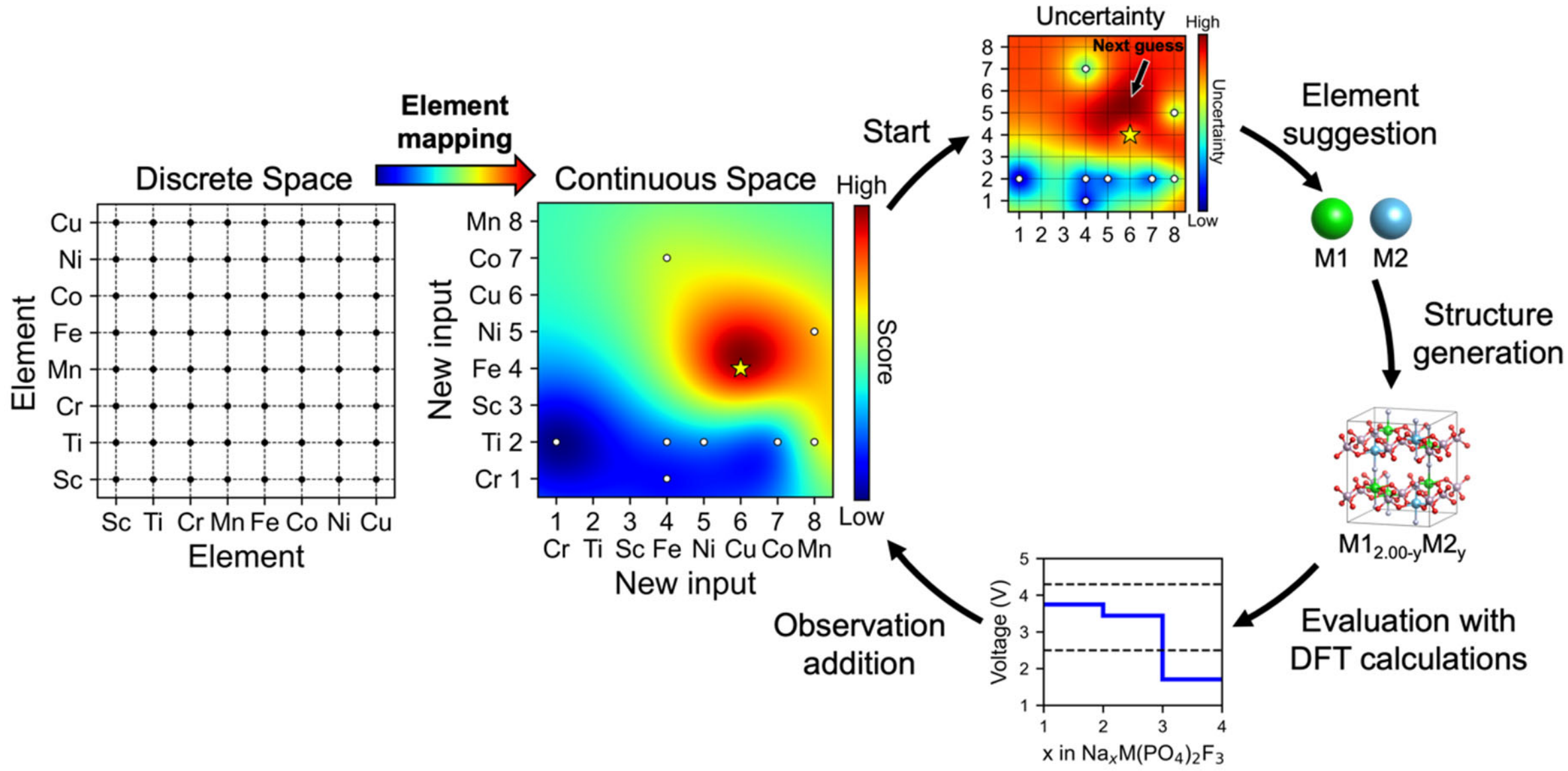

**Fig. 1 | Scheme of Bayesian optimization-based optimal elements discovery algorithm**

## 2. Results & discussion

### 2.1 Algorithm Framework

Fig. 1 illustrates the overall schematic of the Bayesian optimization-based algorithm for discovering optimal binary element combinations that stabilize the sodium-excess phase ($Na_4V_2(PO_4)_2F_3$) of NVPF, thereby repositioning all working voltages within the voltage window of 2.5-4.3V. Although the lattice theoretically allows additional sodium storage, forming a sodium-excess phase, the thermodynamic instability of this sodium-excess phase places its working voltage outside the operational voltage window (2.5–4.3V), thereby causing capacity loss. Stabilizing this sodium-excess phase by confining all redox reactions within the 2.5–4.3 V window could significantly improve the energy density.

Generally, Bayesian Optimization is used to finding the optimal solution of an objective function by predicting with a surrogate model and suggesting the next guess with an acquisition function. However, Gaussian Processes, commonly adopted in surrogate models, are designed for smooth and continuous functions[24], making it challenging to identify optimal chemical components in discrete chemical spaces directly defined by elements. Furthermore, the method for constructing the chemical space is often unclear.

To address this issue, we introduce an element mapping technique. Element mapping involves assigning each element to its continuous value, which transforms the discrete space directly created by the elements into a continuous one. The acquisition function in Bayesian optimization suggests specific values in chemical space as its next guess, and these values are then converted into element combination with element mapping. Based on the proposed element combination, structures are created considering all possible compositions rather than optimizing, since the number of possible cases is small. For example, from the 8 sites for M1 and M2 in the $Na_3M1_{2.00-y}M2_y(PO_4)_2F_3$ (Fig. 2a), the possible chemical compositions replacing vanadium in NVPF are $M1_{1.75}M2_{0.25}$, $M1_{1.50}M2_{0.50}$, $M1_{1.25}M2_{0.75}$, $M1_{1.00}M2_{1.00}$, $M1_{0.75}M2_{1.25}$, $M1_{0.50}M2_{1.50}$ and $M1_{0.25}M2_{1.75}$. The arrangement of the proposed elements within the same composition was analyzed by comparing the relative energies of possible configuration using DFT calculations. The structure with the lowest energy was identified and selected as the representative structure for the proposed chemical composition. From this representative structure, the sodiation process is simulated through DFT calculations to obtain voltage profiles based on the Nernst equation[25]. These voltage profiles are then converted into single values using a scoring function (Fig. 2b and 2c). From this, the highest score from various compositions and components represents the candidates of element combinations. The DFT calculation-based observation is added to the Gaussian process as train data and the algorithm proposes the next guess as a potential element combination. The algorithm iterates this process to find the optimal element combinations with the least number of observations.

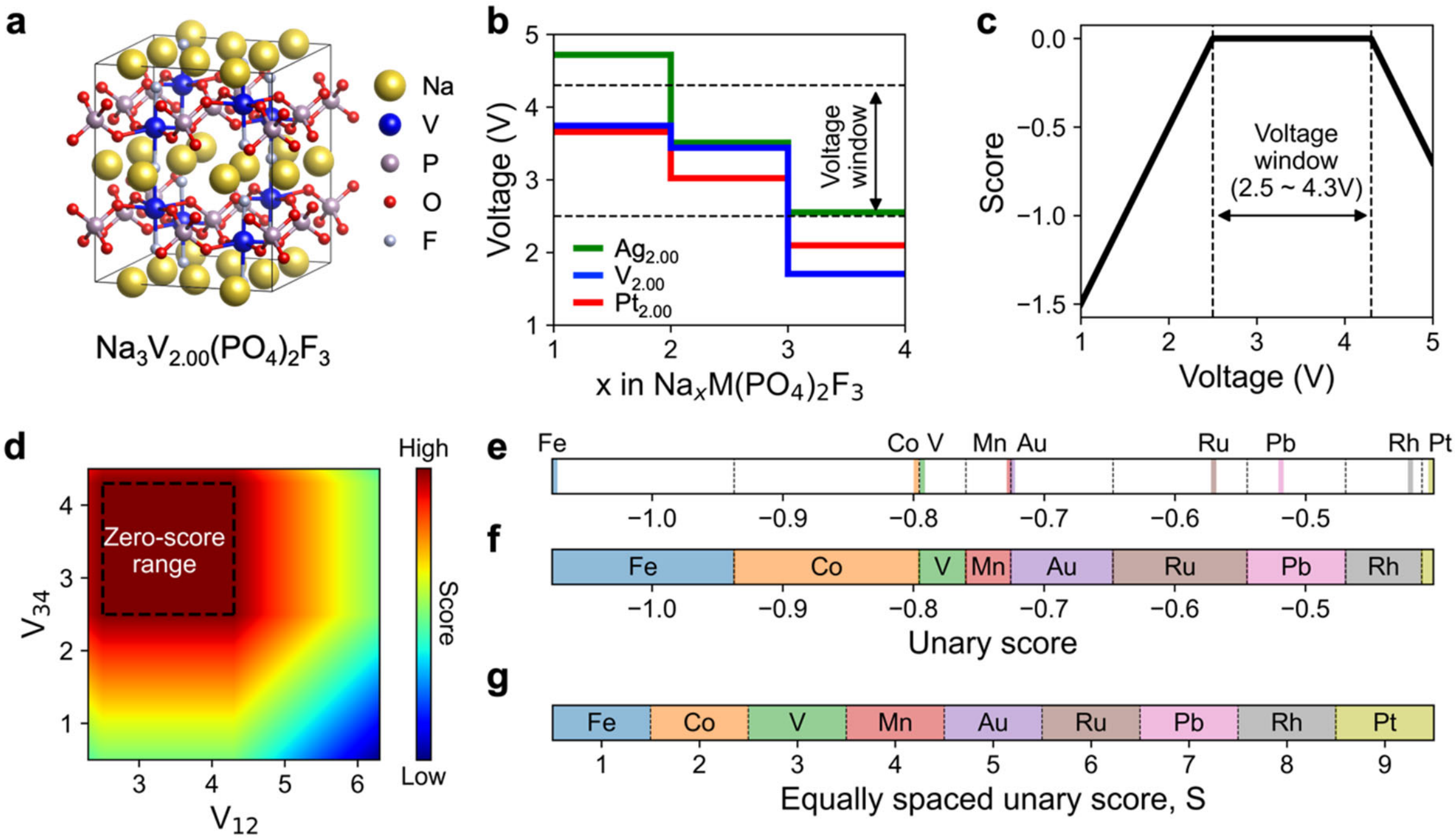

**Fig. 2 | Preprocess for evaluating method and the element mapping. a** Crystal structure of NVPF. **b** DFT calculations-based theoretical voltage profiles of pure NVPF and Ag and Pt substituted NVPF. Scoring function in **c** one-dimension and **d** two-dimension. **e** Unary scores of substituted NVPF with example elements. **f** The regions defined with the median of the unary scores. **g** equally spaced unary score.

## 2.2 Creating chemical space element mapping

Fig. 2 shows details of the scoring function and element mapping. NVPF structures with different sodium concentrations ($Na_1V_2(PO_4)_2F_3$, $Na_2V_2(PO_4)_2F_3$, $Na_3V_2(PO_4)_2F_3$ and $Na_4V_2(PO_4)_2F_3$) are labeled as $Na_1$, $Na_2$, $Na_3$, and $Na_4$ (Fig. 2a and Supplementary Fig. 1). DFT calculations were performed for structures with different sodium concentrations, and theoretical voltage profiles were derived with the Nernst equation (Fig. 2b). The working voltages in the theoretical voltage profiles for the ($Na_1$, $Na_2$), ($Na_2$, $Na_3$), and ($Na_3$, $Na_4$) structures are specified as $V_{12}$, $V_{23}$, and $V_{34}$. To create a criterion for the Bayesian optimization to evaluate voltage profiles from suggested element combination, we define a scoring function (Fig. 2c). The scoring function gives a penalty when the working voltage deviates from the voltage window. When the working voltage is within the voltage window the scoring function returns zero as a highest score. The total score of the voltage profile is defined as the sum of the scoring function values of $V_{12}$, $V_{23}$, and $V_{34}$.

Fig. 2d shows the scoring function in two-dimension constructed with $V_{12}$ and $V_{34}$. The black dash line box in Fig. 2d indicates the zero-score range area, where total score equals zero. In the practical optimization process, a three-dimensional scoring function is formed with score function about $V_{12}$, $V_{23}$ and $V_{34}$. This scoring method aims to identify the optimal element combination that brings all working voltages of the structure within the zero-score range. To successfully find the optimal element combinations with the minimum number of observations, the chemical space defined by the input and the scoring function (output) should be easy to predict. This means that there should be fewer regions with steep gradients within the chemical space. To create easy to predict chemical space, the input and the scoring function should be highly correlated to avoid abrupt gradient regions. Because the scoring function evaluates the position of working voltage, the scoring function is highly connected to the sodiation process. For this reason, the input should be correlated with the sodiation process.

Here, we introduce a unary score as an appropriate input. The unary score represents the evaluation of a structure in which all vanadium atoms in NVPF are substituted with the specified element. From this, it is possible to capture the properties of the sodiation process as a value when the element is within the framework of an NVPF. For example, the theoretical voltage profiles of pristine NVPF and the fully substituted NVPF with Pt and Ag are shown in Fig. 2b, and the order of their unary score is Pt, V and Ag (-0.40, -0.79 and -2.01, respectively). In Fig. 2e, the unary scores of nine example elements are described in one-dimensional continuous space. Within the continuous space, the regions are defined with the median of the unary scores as the boundary (Fig. 2f). By equally spacing these regions (Fig. 2g), the equally spaced regions are assigned to the elements and the elements can be mapped to the specific value.

Even if not the unary score, any value that can represent an element can be used for element mapping. One example is electronegativity, which represents the tendency to attract electrons and is already used as an input in machine learning models[26-28]. Because the role of transition metals in NMPF is to store electrons during the sodiation process, electronegativity can be useful for element mapping. However, the unary score has a higher correlation with the scoring function than electronegativity. The unary score provides a direct assessment of the role for transition metal within the NVPF environment, capturing aspects beyond what electronegativity alone can reveal. It also accounts for relaxation effects and alterations in the electronic structure. To investigate how the correlation between the sodiation process and input facilitates the prediction of chemical space, 35 elements were mapped using three equally spaced parameters: the unary score (S), electronegativity (X), and atomic number (Z), to serve as inputs (Supplementary Fig. 2-4).

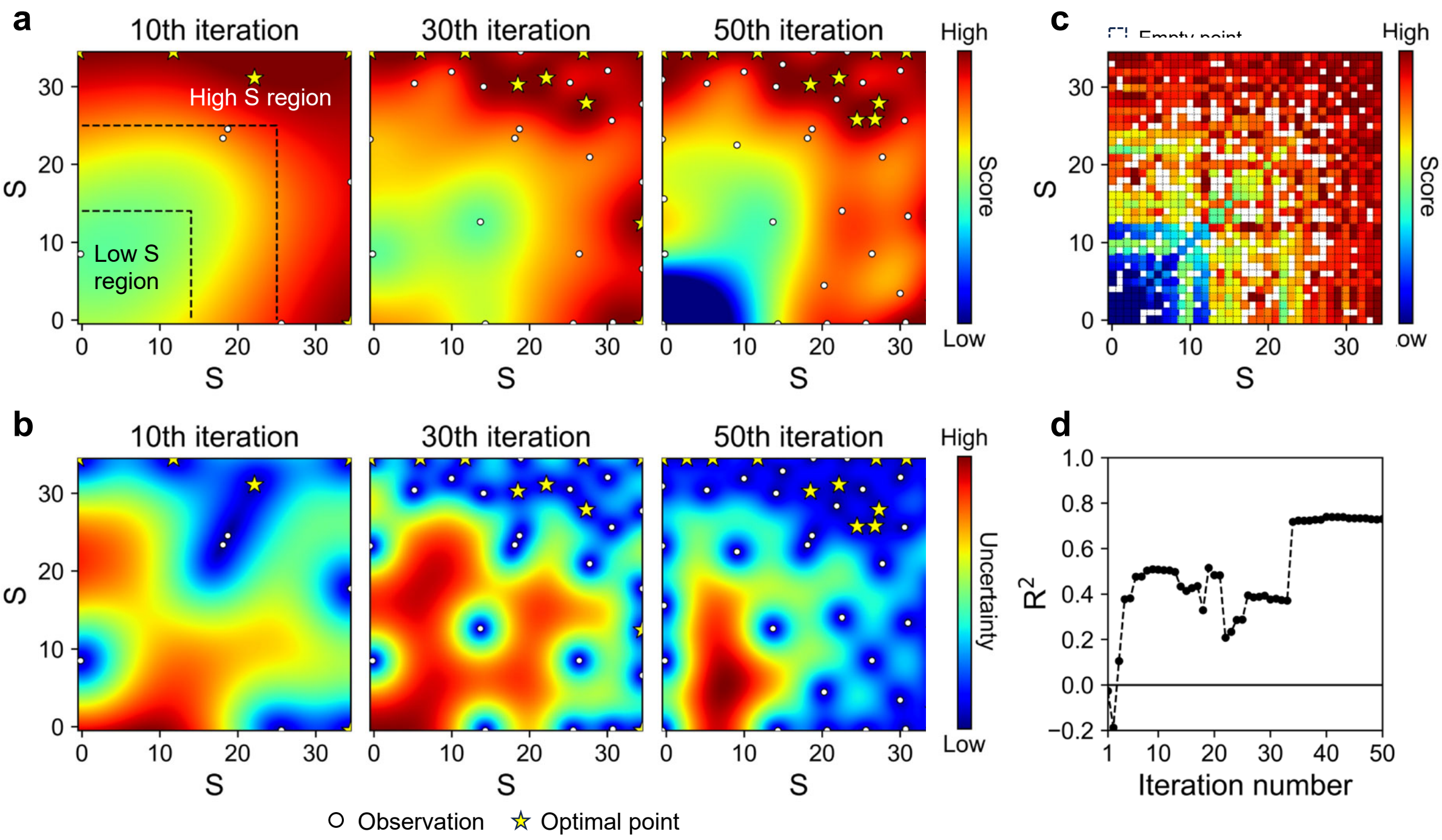

**Fig. 3 | The process and results of exploring binary element combination. a** Predicted chemical space at 10th, 30th and 50th iterations. **b** Uncertainty at 10th, 30th, 50th iterations. White circle and yellow star indicate observation and optimal point. The predicted score and uncertainty are represented by colors. **c** Ground truth defined by the collected observation data. White area indicates empty point. **d** $R^2$ of predicted chemical space at each iteration.

| Input | S | | X | Z |
|---|---|---|---|---|
| Optimal composition ($M1_{2.00-y}M2_y$) | $Pd_{1.75}Re_{0.25}$ | $Co_{1.00}Ir_{1.00}$ | $Pd_{1.75}Re_{0.25}$ | $Pd_{1.75}Re_{0.25}$ |
| | $Pd_{1.50}Re_{0.50}$ | $Co_{0.75}Ir_{1.25}$ | $Pd_{1.50}Re_{0.50}$ | $Pd_{1.50}Re_{0.50}$ |
| | $Pd_{1.50}Ta_{0.50}$ | $Co_{0.50}V_{1.50}$ | $Ir_{1.00}Pb_{1.00}$ | $Au_{1.25}Mo_{0.75}$ |
| | $Pd_{1.25}V_{0.75}$ | $Os_{0.75}Pb_{1.25}$ | $Ir_{0.75}Pb_{1.25}$ | $Au_{1.25}Ta_{0.75}$ |
| | $Pd_{1.75}W_{0.25}$ | $Os_{0.50}Pb_{1.50}$ | $Mo_{0.75}Tl_{1.25}$ | $Mo_{0.75}Tl_{1.25}$ |
| | $Pb_{0.50}Pd_{1.50}$ | $Mn_{0.75}V_{1.25}$ | $Au_{1.25}Nb_{0.75}$ | $Os_{0.75}Pb_{1.25}$ |
| | $Pb_{0.75}Pd_{1.25}$ | | $Au_{1.25}Mo_{0.75}$ | $Os_{0.50}Pb_{1.50}$ |
| | $Pb_{1.50}Pd_{0.50}$ | | $Pb_{0.75}Pt_{1.25}$ | |
| | $Hf_{0.50}Pd_{1.50}$ | | $Mn_{0.50}Pb_{1.50}$ | |
| | $Mo_{0.50}Pd_{1.50}$ | | | |
| Total | 16 | | 9 | 6 |

**Table. 1 | Discovered optimal element composition.**

## 2.3 Bayesian optimization for optimal binary element combination

Fig. 3 shows the process and results of exploring the optimal binary element combinations in a two-dimensional continuous chemical space transformed with element mapping by equally spaced unary score (S). The chemical space is formed with 35 elements and the total number of cases is 630 ($_{35}C_1 + _{35}C_2$). The predicted chemical spaces by the surrogate model at the 10th, 30th, and 50th iterations are shown in Fig. 3a. In the predicted chemical spaced at the 10th iteration steps, there is a preliminary tendency for low S regions to have low scores and high S regions to have high scores. As the iterations progress, the trend becomes clearer, and it has been found that there are more observations in the high S region with high scores rather than in the low S region. It means that the exploitation to observe near the high score observations progressed properly. Fig. 3b shows the uncertainty from the 95% confidence interval at the 10th, 30th, and 50th iterations. The maximum and minimum values of uncertainty are determined in each iteration step. As the iterations progress, uncertainty within the chemical space has been reduced overall. Specifically, at the 50th iteration, the uncertainty of low S region is still high and there are relatively few observations. It means that the algorithm has primarily explored high S regions, demonstrating exploitation by focusing on areas with high scores. Also, given that there are observations in the 50th iteration in areas where the algorithm already judged the score to be low at the 30th iteration, it is inferred that there has been progress in exploration as well as exploitation. In case of optimization with element mapping by S, 16 optimal points are discovered, and 14 element combinations are discovered without overlapped element combinations (Os-Pb, Pd, Hf-Pd, Mo-Pd, Pd-V, Mn-V, Pd-Re, Mo-Pd, Ag-Ru, Pd-W, Co-Ir, Pb-Pd, Co-V and Pd-Ta). A total of 24 compositions are included in element combinations. For validation, DFT calculations with tight convergence criteria for voltage profiles of 24 compositions are performed again. Finally, 16 compositions are confirmed that all working voltages are within voltage window (Table. 1 and Supplementary Fig. 5).

To evaluate the accuracy of surrogate model, the ground truth is defined with the obtained observation data (Fig. 3c). Unobserved data are considered as empty points. A qualitative comparison of the predicted chemical space at iteration 50 with the ground truth shows a similar trend of higher scores in the rightward direction. Additionally, by measuring the $R^2$ between predicted chemical space and ground truth per iteration step, the quantitative accuracy was identified (Fig. 3g). The $R^2$ increases until about 10th iteration and then gradually decreases to the 35th iteration. This is due to the lack of observations in the low S region at 30th iteration. This makes large difference in low S region of predicted chemical space and ground truth. However, after the exploratory observations in the low S region, $R^2$ is suddenly increased and high $R^2$ is achieved (0.730 at 50th iteration).

To understand the differences in constructing the chemical space with the different inputs, the two-dimensional chemical spaces were constructed with X and Z. Supplementary Figs. 6a and 7a show the predicted two-dimensional space corresponding to each iteration step by the surrogate model and Supplementary Figs. 6b and 7b show the uncertainty. For each case, comparing the predicted chemical space at 50th iteration with their ground truth (supplementary Figs. 6c and 7c) shows that there are a lot of differences. In particular, for X, despite the overall low uncertainty at the 50th iteration, there are the large difference with the ground truth indicating that the model misunderstands the trend of the chemical space. In supplementary Fig. 6d, $R^2$ stays around -0.25 until the 20th iteration, after which it decreases sharply, becoming very low near the 30th iteration and increasing later. However, it still has negative $R^2$ at the 50th iteration. Also, in the case of Z, $R^2$ remains around 0.15 and does not increase with iterations (Supplementary Fig. 7d). These mean that the surrogate model doesn't capture the trends in the chemical space correctly, resulting in predictions that deviate significantly from the ground truth.

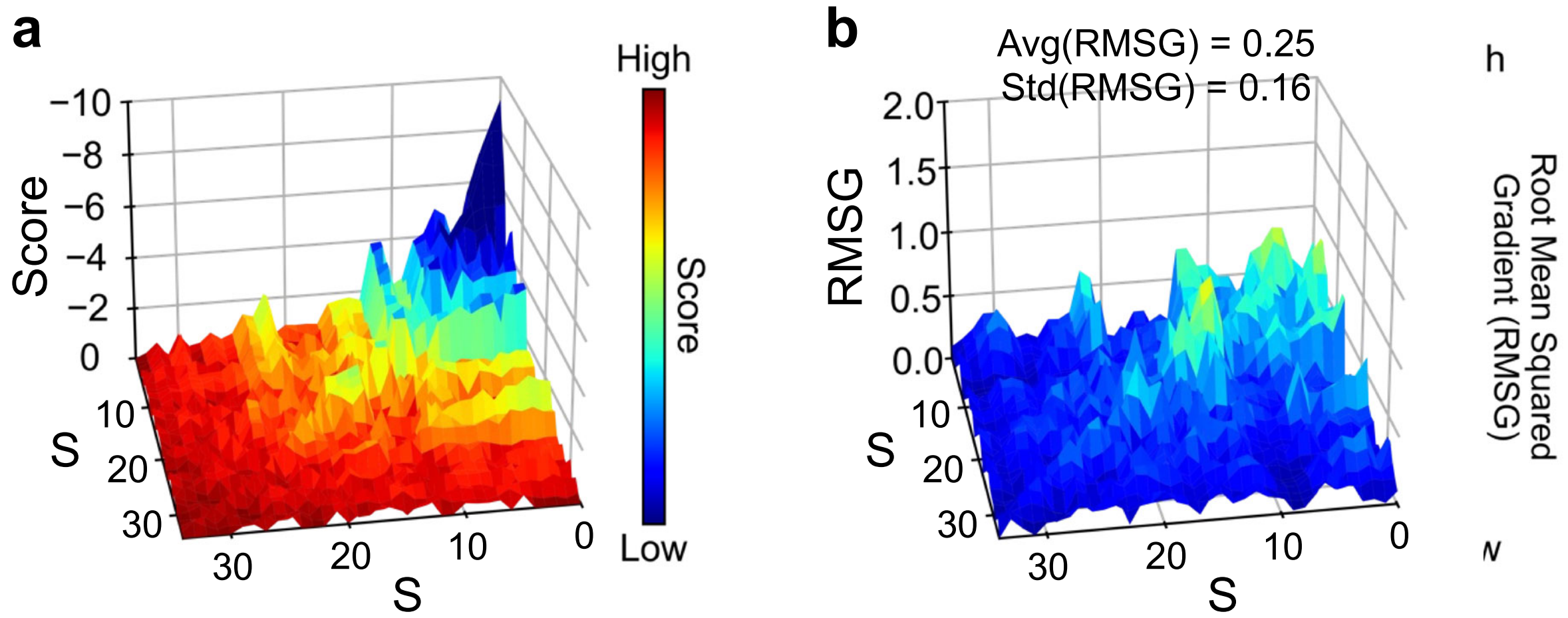

**Fig. 4 | The morphology and magnitude of gradient of chemical space. a** The three-dimensional map of chemical space constructed with S. **b** Root mean squared gradient (RMSG) of score in chemical space. The score and RMSG are represented by colors.

The difference in the number of discovered optimal points and accuracy is mainly due to the difficulty of the constructed chemical space. Because the Gaussian process is designed for smooth and continuous functions, it is hard to predict highly rough functions. Fig. 4a, and supplementary Figs. 8a and 8c show the morphology of chemical space. The empty points in chemical space are filled with average value of their neighbors. The morphology of chemical space formed with S (Fig. 4a) shows qualitatively low complexity and high smoothness rather than others (supplementary Figs. 8a and 8c). Specifically, in the chemical space constructed with S, there are fewer sharp differences in scores in neighboring areas. Also, the high score points (optimal points) are clustered. However, the chemical space formed with X and Z shows drastic difference in scores in neighboring area and their morphology is more complex. To analyze the difference quantitatively, we defined and visualized the root mean squared gradient (RMSG) with below equation, which quantifies the gradient between the point and their neighbors in each chemical space (Fig. 4b, supplementary Figs. 8b and 8d).

$$RMSG_{i,j} = \frac{1}{N}\sqrt{(M_{i,j} - M_{i,-1\,j})^2 + (M_{i,j} - M_{i,+1\,j})^2 + (M_{i,j} - M_{i,\,j-1})^2 + (M_{i,j} - M_{i,\,j+1})^2}$$

Here, $M_{i,j}$ refers to the value at $(i,j)$ position within the matrix (chemical space), while $N$ is the number of neighbors. In Fig. 4b, there are fewer large gradients than others (Supplementary Figs. 8b and 8d). Also, in case of S, the average and standard deviation of RMSG (0.25 and 0.16) is most small than others (0.48 and 0.39 for X and 0.34 and 0.26 for Z). It means that the chemical space with S is most smooth and easy to predict. The difference in difficulty is caused by the correlation between the output of the chemcial space, the scoring function, and the input. Specifically, the scoring function is highly related with sodiation. Considering this, we can infer that because unary score is so closely related to sodiation, it has been able to construct a chemical space with the certain tendency which makes smooth chemical space.

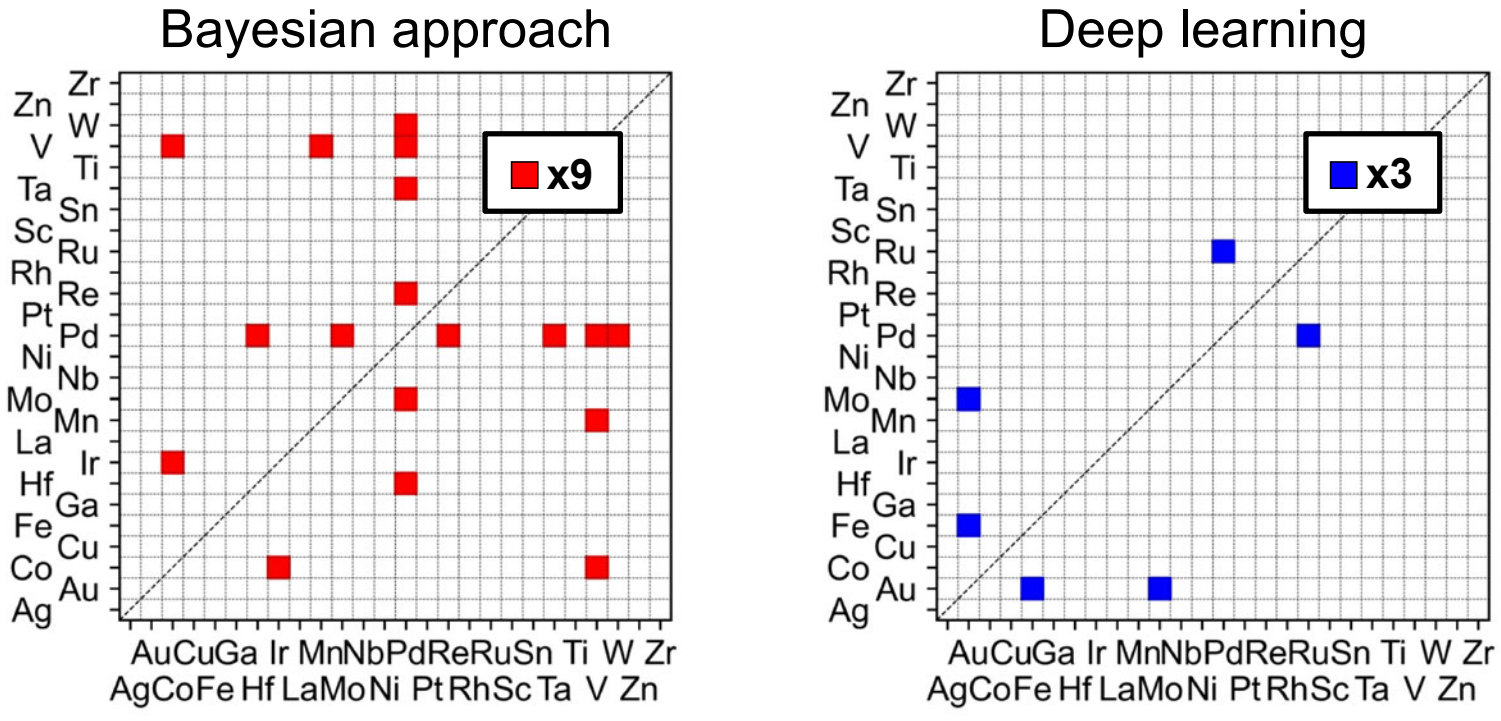

**Fig. 5 | Schematic comparison of Bayesian approach and deep learning-based approach**

## 2.4 Effectiveness of the Proposed Algorithm

To confirm how the new application of Bayesian optimization is effective in finding optimal element combinations, we compared our results with those of deep learning-based screening study[5]. In the previous study, the deep learning model[4] predicts the formation energy of $Na_xM1_{2.00-y}M2_y(PO_4)_2F_3$ (NMPF) in a chemical space formed with 30 elements (Supplementary Fig. 8b) where the total number of cases is 465 ($_{30}C_1$ + $_{30}C_2$). Because the candidate elements for constructing chemical space are different, we only consider overlapping elements in the candidate elements of each study (Fig. 5). For previous study, the predicted working voltages of NMPF ($V_{12}$, $V_{23}$ and $V_{34}$) based on the predicted formation energy are adapted to the scoring function defined in this study (Figs. 2b and 2c) and a total of 27 predicted optimal points are found. However, after the validation process with DFT calculations with the tight convergence criteria, there are only four optimal binary compositions ($Au_{1.25}Mo_{0.75}$, $Au_{1.50}Fe_{0.50}$, $Pd_{1.25}Ru_{0.75}$ and $Pd_{1.50}Ru_{0.50}$) and three element combinations (Au-Mo, Au-Fe and Pd-Ru). This result shows that the proposed algorithm is more effective in terms of finding more optima. The difference in the number and kinds of discovered optimal points is due to the different mechanisms for finding optima in each approach. The approach of previous work observes all points in grid with predictions from machine learning model. This approach has the advantage of quickly exploring the entire chemical space. However, due to an error of machine learning model-based prediction the number of optimal points is reduced after validation in the DFT-level. In contrast, Bayesian optimization finds the optima by suggesting the next guess based on previous DFT-level observations obtained through direct DFT calculations. This allows it to find as many optimal points as possible within a given number of observations. However, since it proceeds from a given number of observations (e.g., 50 iterations out of a total of 630 points), there is a limit to finding all the optimal points in the chemical space. And this limitation is expected to be addressed by conducting several Bayesian optimizations.

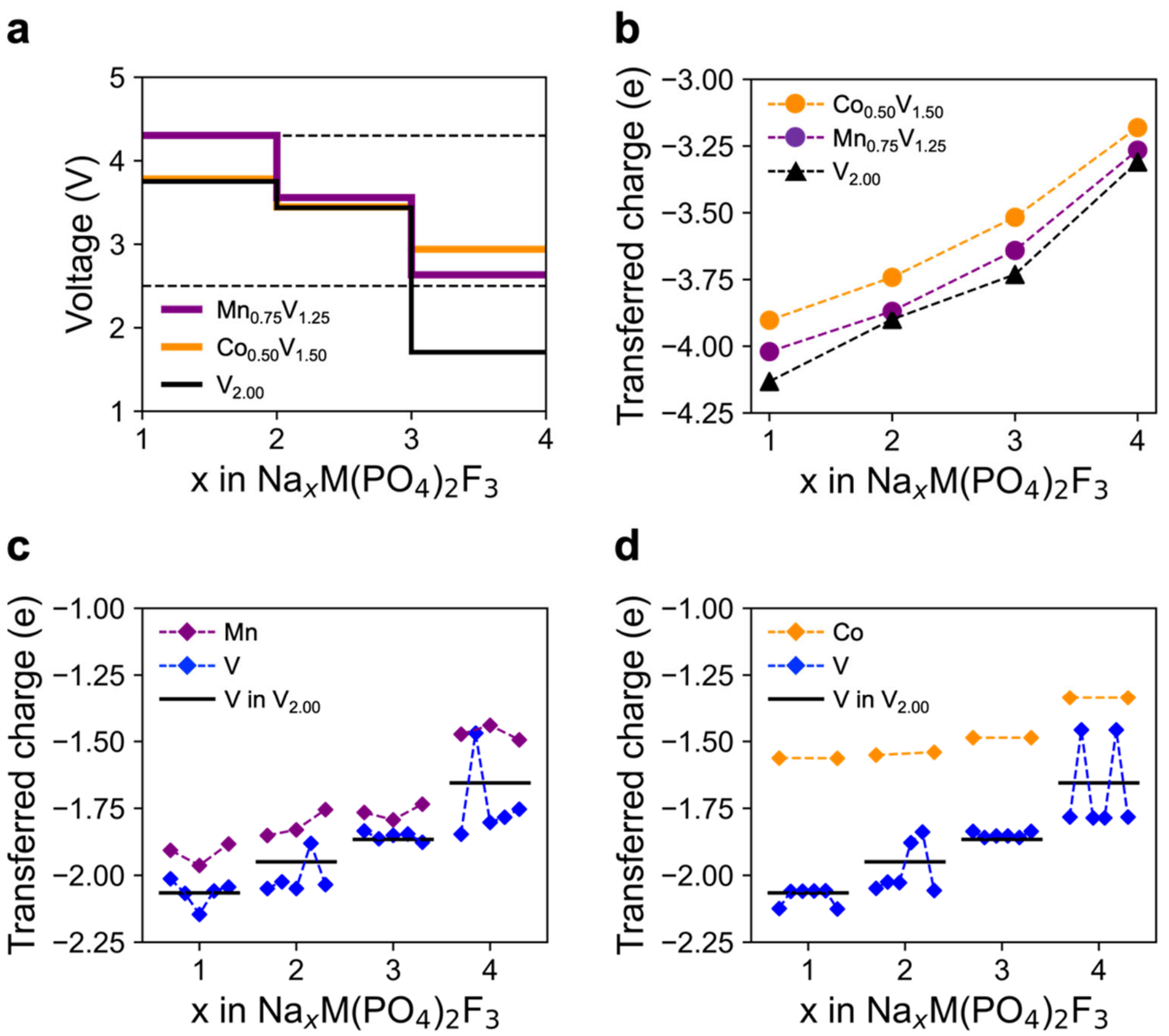

**Fig. 6 | Analysis with DFT Calculations. a** Theoretical voltage profiles of NMPF with $Mn_{0.75}V_{1.25}$ and $Co_{0.50}V_{1.50}$ and NVPF. **b** Bader charge analysis for transferred charge to $Mn_{0.75}V_{1.25}$ and $Co_{0.50}V_{1.50}$ in NMPF and $V_{2.00}$ in NVPF. Atomic Bader charge analysis for transferred charge to **c** $Mn_{0.75}V_{1.25}$ and **d** $Co_{0.50}V_{1.50}$ in NMPF.

## 2.5 Analysis with DFT Calculations

To understand how the discovered optimal element combinations work in NVMF, their electronic properties were analyzed using DFT calculations. Considering the theoretical energy density, cost-effectiveness and toxicity, $Mn_{0.75}V_{1.25}$ and $Co_{0.50}V_{1.50}$ are suitable element combinations (Fig. 6a and Supplementary Fig. 9). To determine the degree of electron acceptance on the NMPF according to the sodium concentration, Bader charge analysis[29], which can analyze the degree of the charge transfer, is performed. Fig. 6b and Supplementary Fig. 10a-c show the total amount of charge transferred to the element combination for discovered NMPF. In all cases, the total amount of charge transferred to the element combination in NMPF is higher than the transferred charge of the NVPF ($V_{2.00}$). This implies that one of the conditions for working voltages to be within the voltage window is to accept more electrons than vanadium during sodiation. Specifically, Fig. 6b, c and Supplementary Fig. 11 show atomic Bader charge analysis of the NMPF with element combinations containing vanadium ($Mn_{0.75}V_{1.25}$, $Co_{0.50}V_{1.50}$ and $Pd_{1.25}V_{0.75}$). The rhombus represents the charge transferred to each transition metal, and the black line is the average value of the transferred charge of V in NVPF ($V_{2.00}$). The amount of charge transferred to vanadium in the $Na_1$, $Na_2$, and $Na_3$ phases is similar to the transferred charge of pure NVPF (V in $V_{2.00}$), but there is a different in the $Na_4$ phase. This suggests that the formation of the $Na_4$ (sodium excess phase) is unstable due to a lot of transferred charge to vanadium, but the NMPF with optimal element is energetically stable by transferring the charge to a substituted transition metal that can more readily accept the charge rather than vanadium.

## 3. Conclusions

In this study, we propose an algorithm to discover element combinations that address the limitation of the Gaussian process in Bayesian optimization. By assigning elements to continuous values through element mapping, it is possible to construct a chemical space for applying Bayesian optimization.

The main advantage of this proposed algorithm is its capability to effectively navigate low-dimensional chemical spaces with limited observations. Unlike deep learning-based grid search methods that require extensive training data and computational resources for training, our approach does not require initial training data and resources for training. It is also flexible and scalable to a wide range of material systems by defining the scoring function and input appropriately.

Through its application, we identified 16 optimal combinations of elements in just 50 iterations and designed high-capacity NASICON-type cathode materials for sodium-ion batteries—specifically $Mn_{0.75}V_{1.25}$ and $Co_{0.50}V_{1.50}$—which successfully stabilize the sodium-excess phase of NVPF.

This new framework presents a robust, extensible, and flexible platform for data-driven material design by combining Bayesian optimization and element mapping. It uses chemically intuitive unary scores—continuous descriptors derived from DFT—to bypass large training datasets and enable mechanistic interpretability. As a result, it drives rapid convergence in discrete chemical spaces and accelerates discovery across diverse material classes, opening new avenues for future data-driven discovery.

**Acknowledgements**

This research was supported by the Nano & Material Technology Development Program through the National Research Foundation of Korea (NRF) funded by Ministry of Science and ICT (RS-2024-00449682) and Yangyoung Foundation in 2025.

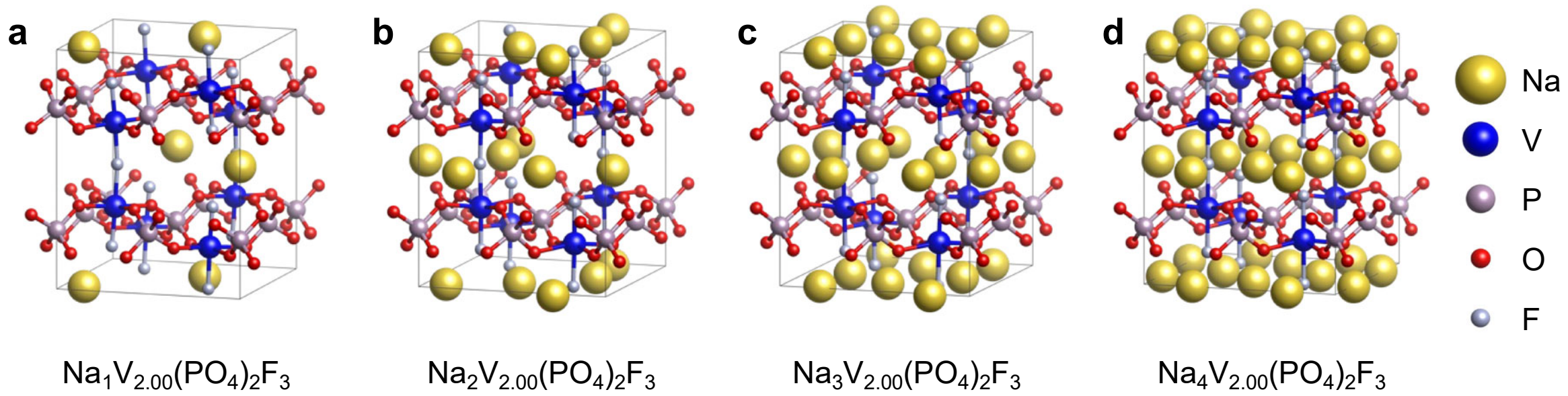

**Supplementary Fig. 1 | The crystal structure of NVPF with different sodium concentration. a** $Na_1$, **b** $Na_2$, **c** $Na_3$ and **d** $Na_4$.

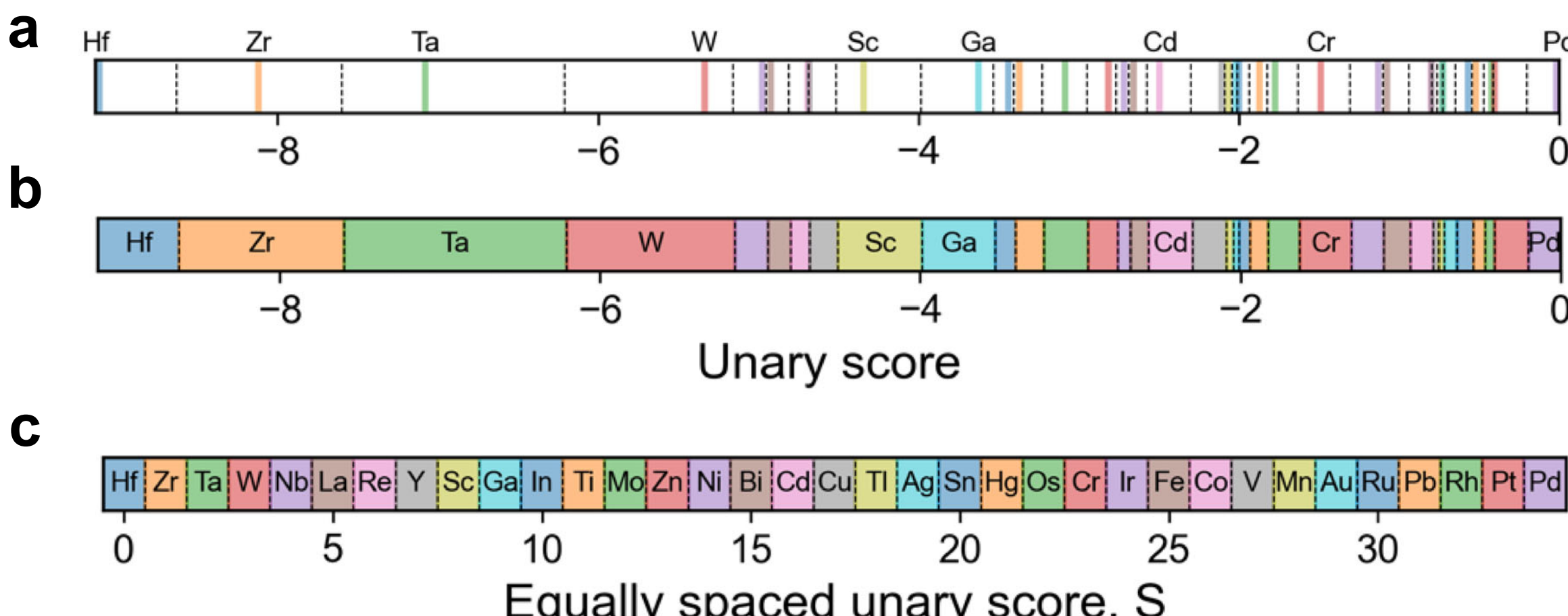

**Supplementary Fig. 2 | Definition of input with element mapping. a** Unary scores of substituted NVPF with 35 candidate elements. **b** The regions defined with the median of the unary scores. **c** Equally spaced unary score.

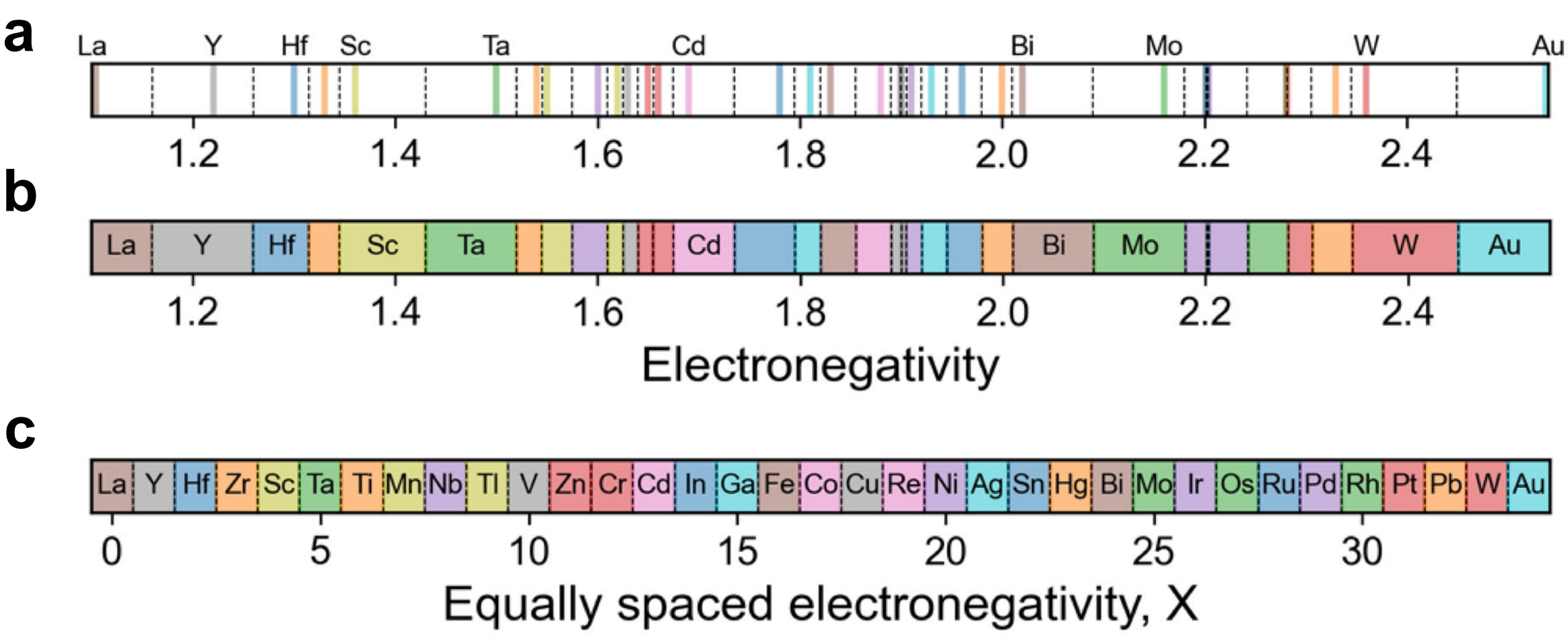

**Supplementary Fig. 3 | Definition of input with element mapping. a** Electronegativities of 35 candidate elements. **b** The regions defined with the median of the electronegativities. **c** Equally spaced electronegativity.

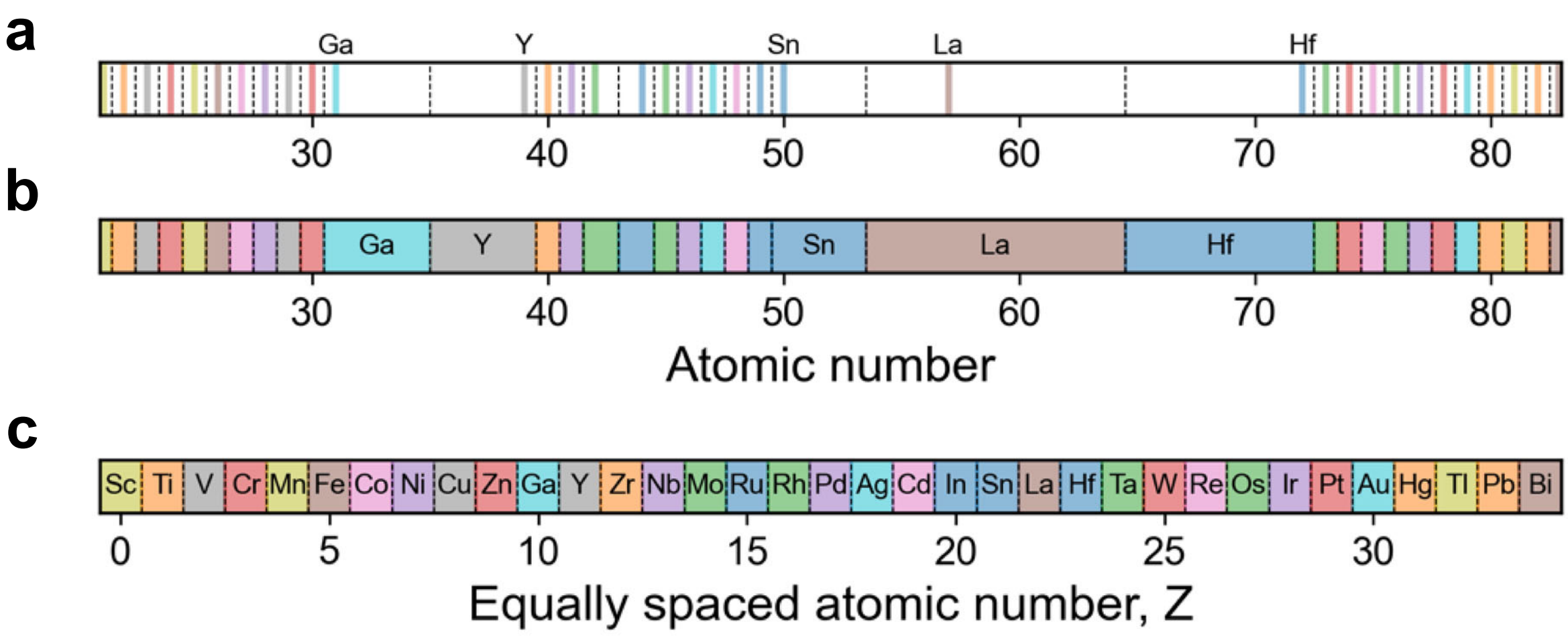

**Supplementary Fig. 4 | Definition of input with element mapping. a** Atomic numbers of 35 candidate elements. **b** The regions defined with the median of the atomic numbers. **c** Equally spaced atomic number.

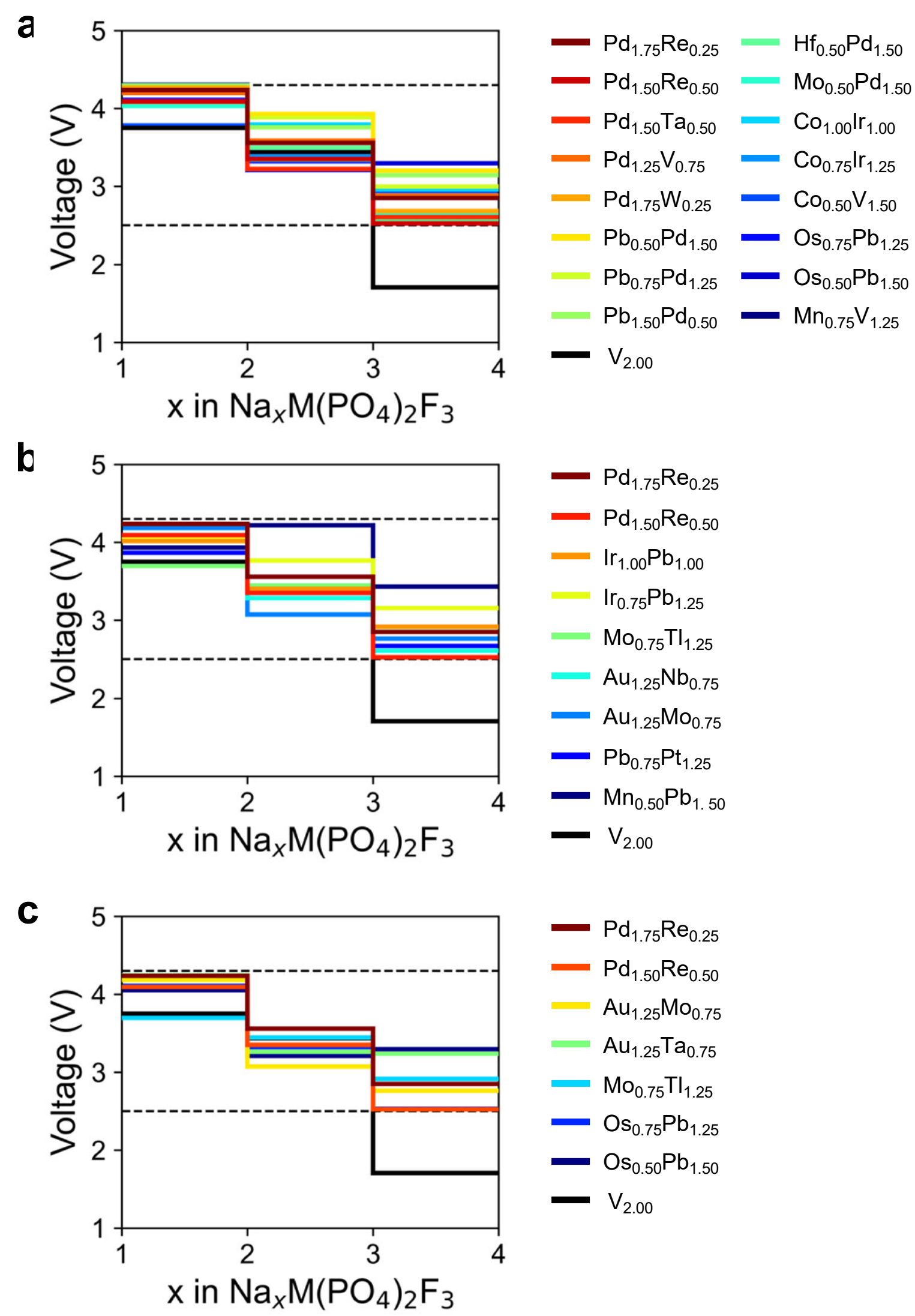

**Supplementary Fig. 5 | The theoretical voltage profiles.** The theoretical voltage profiles of NMPF with optimal binary compositions from chemical space constructed from **a** S, **b** X and **c** Z.

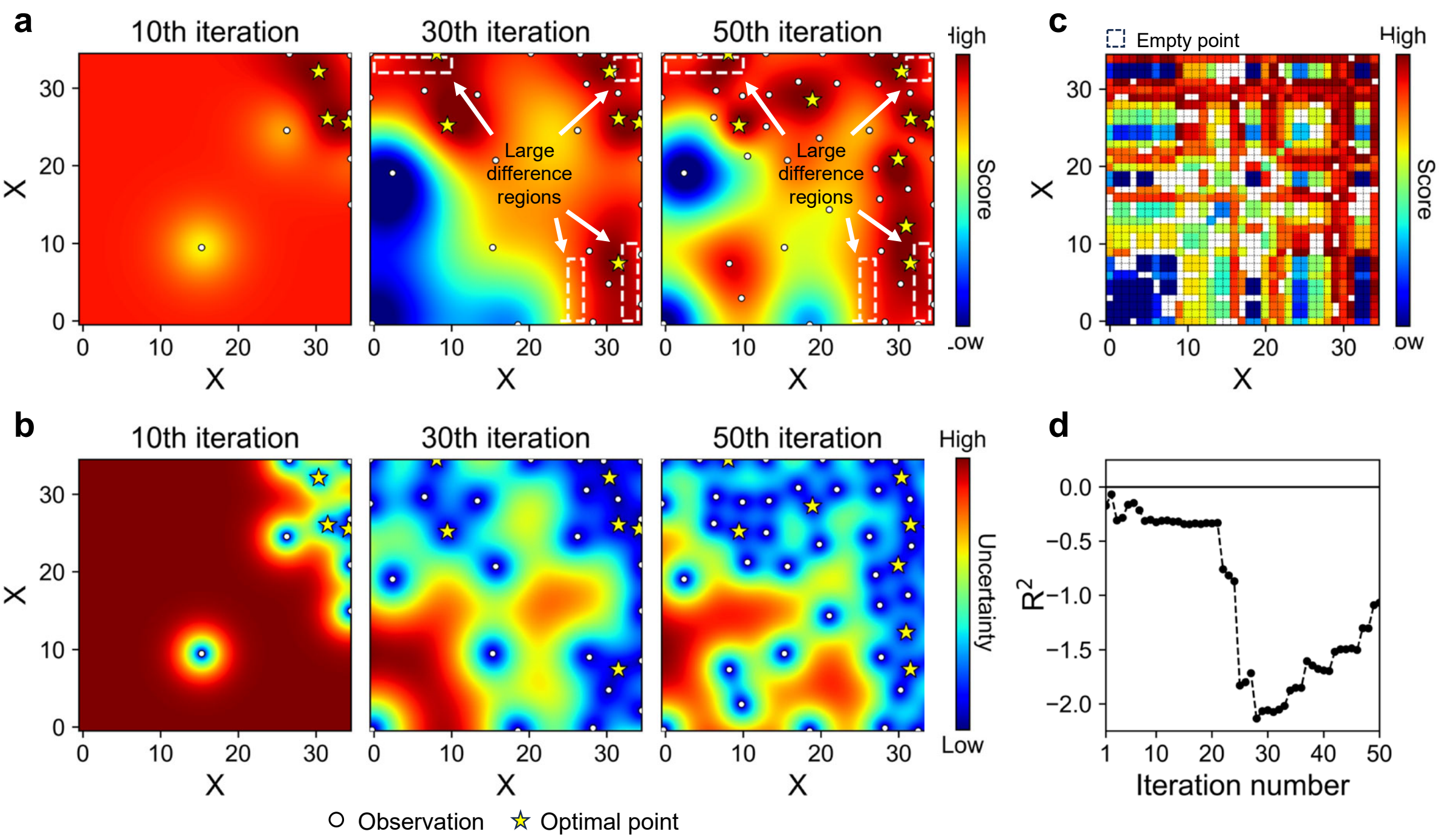

**Supplementary Fig. 6 | The process and results of exploring binary element combination.** Predicted chemical space at **a** 10th, **b** 30th and **c** 50th iterations. Uncertainty at **d** 10th, **e** 30th, **f** 50th iterations. White circle and yellow star indicate observation and optimal point. The predicted score and uncertainty are represented by colors. **g** $R^2$ of predicted chemical space at each iteration. **h** Ground truth defined by the collected observation data. White area indicates empty point. **i** The morphology of constructed chemical space.

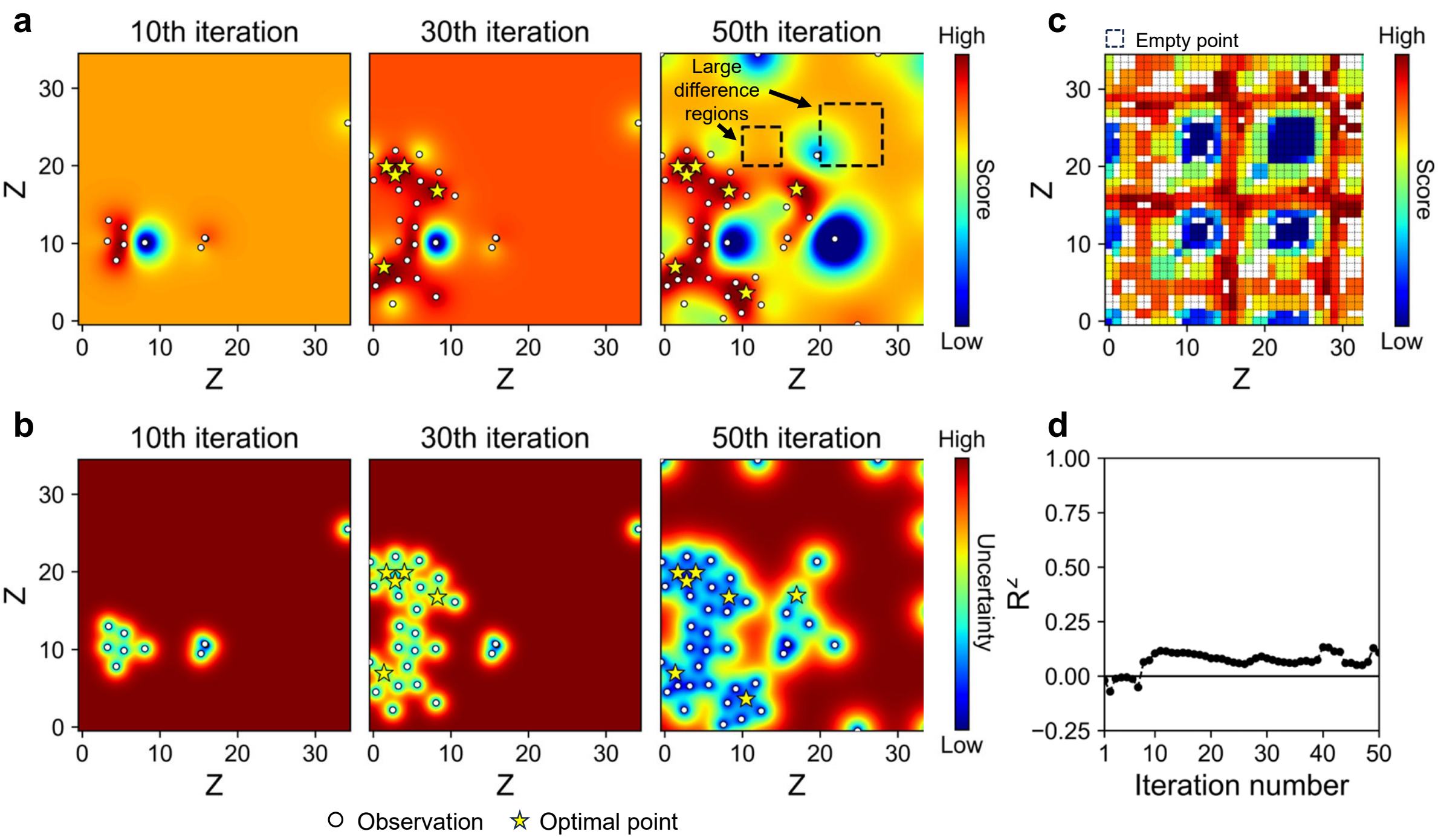

**Supplementary Fig. 7 | The process and results of exploring binary element combination.** Predicted chemical space at **a** 10th, **b** 30th and **c** 50th iterations. Uncertainty at **d** 10th, **e** 30th, **f** 50th iterations. White circle and yellow star indicate observation and optimal point. The predicted score and uncertainty are represented by colors. **g** $R^2$ of predicted chemical space at each iteration. **h** Ground truth defined by the collected observation data. White area indicates empty point. **i** The morphology of constructed chemical space.

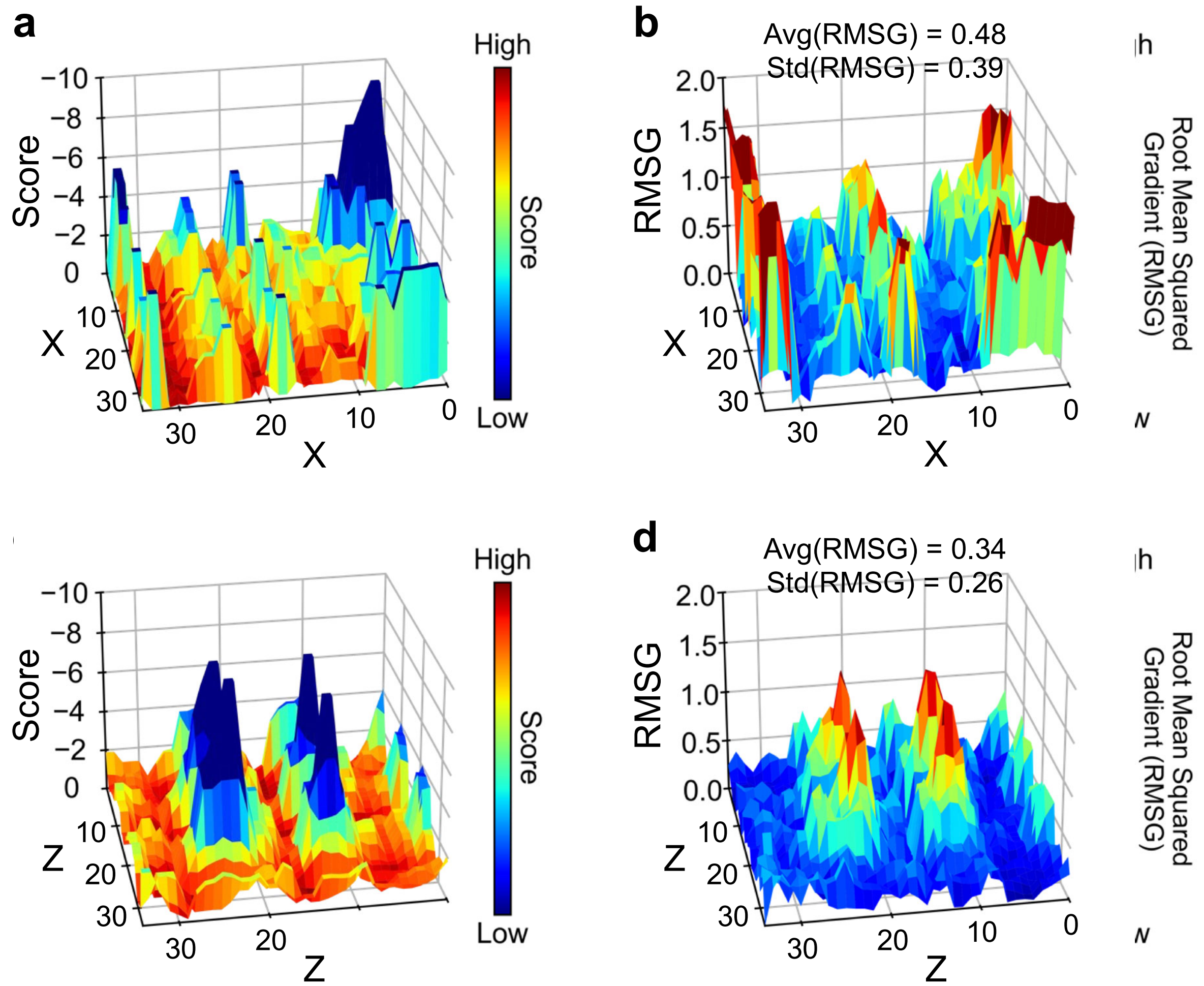

**Supplementary Fig. 8 | The morphology and magnitude of gradient of chemical space.** The three-dimensional map of chemical space constructed with **a** X and **c** Z. Root mean squared gradient (RMSG) of score in chemical space with **b** X and **d** Z. The score and RMSG are represented by colors.

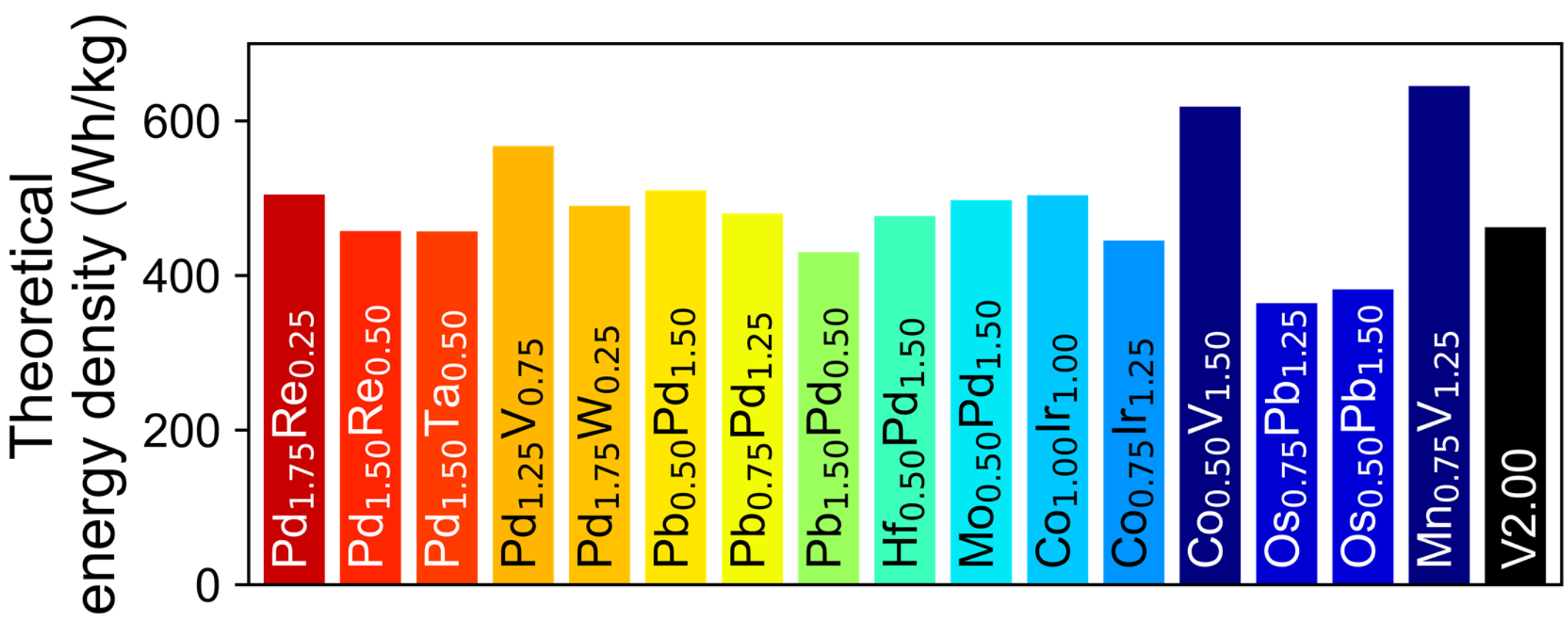

**Supplementary Fig. 9 | The theoretical energy densities of NMPF with optimal binary compositions.**

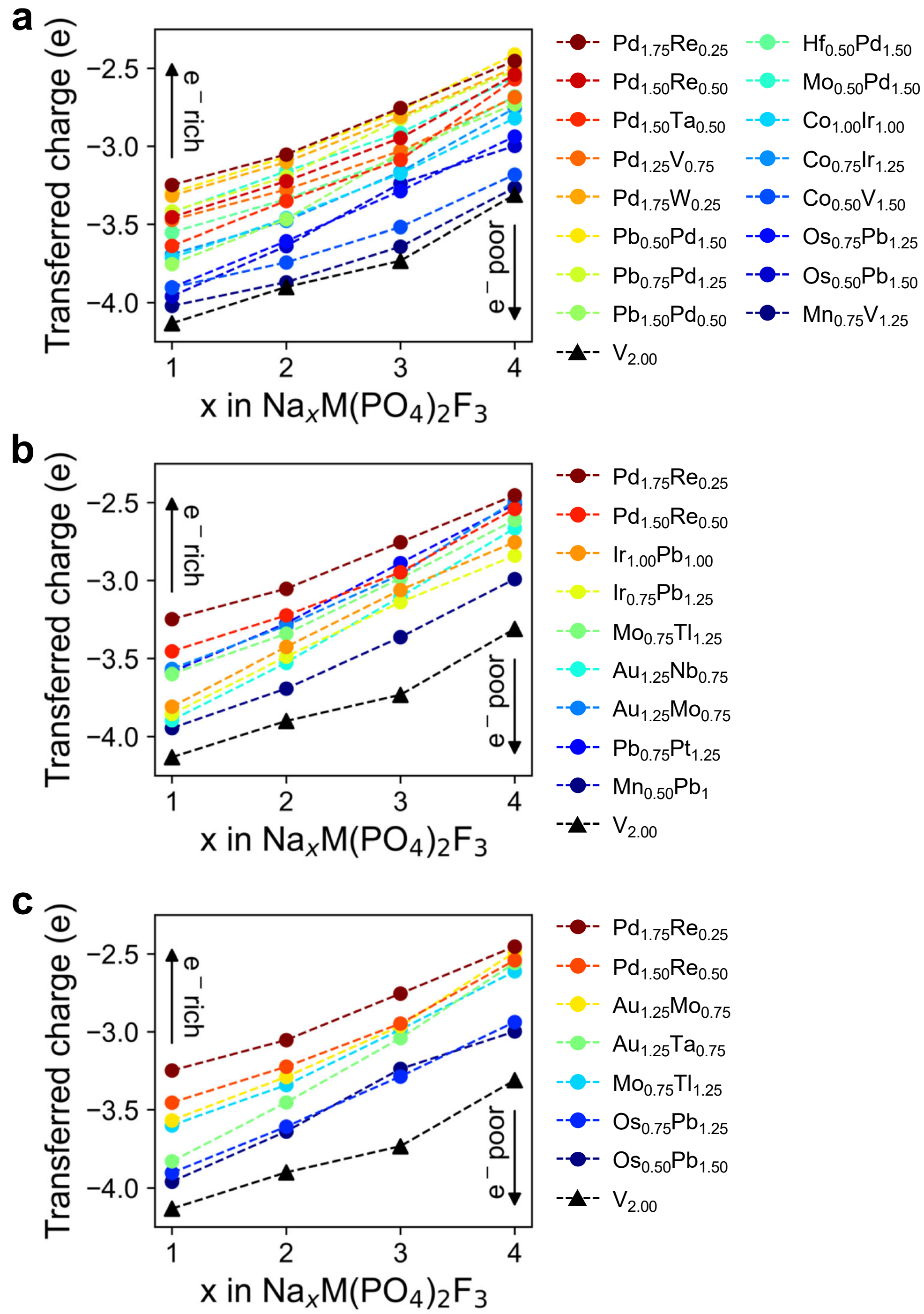

**Supplementary Fig. 10 | Bader charge analysis for transferred charge to optimal binary compositions.** The Bader charge analysis of NMPF with optimal binary compositions from chemical space constructed from **a** S, **b** X and **c** Z.

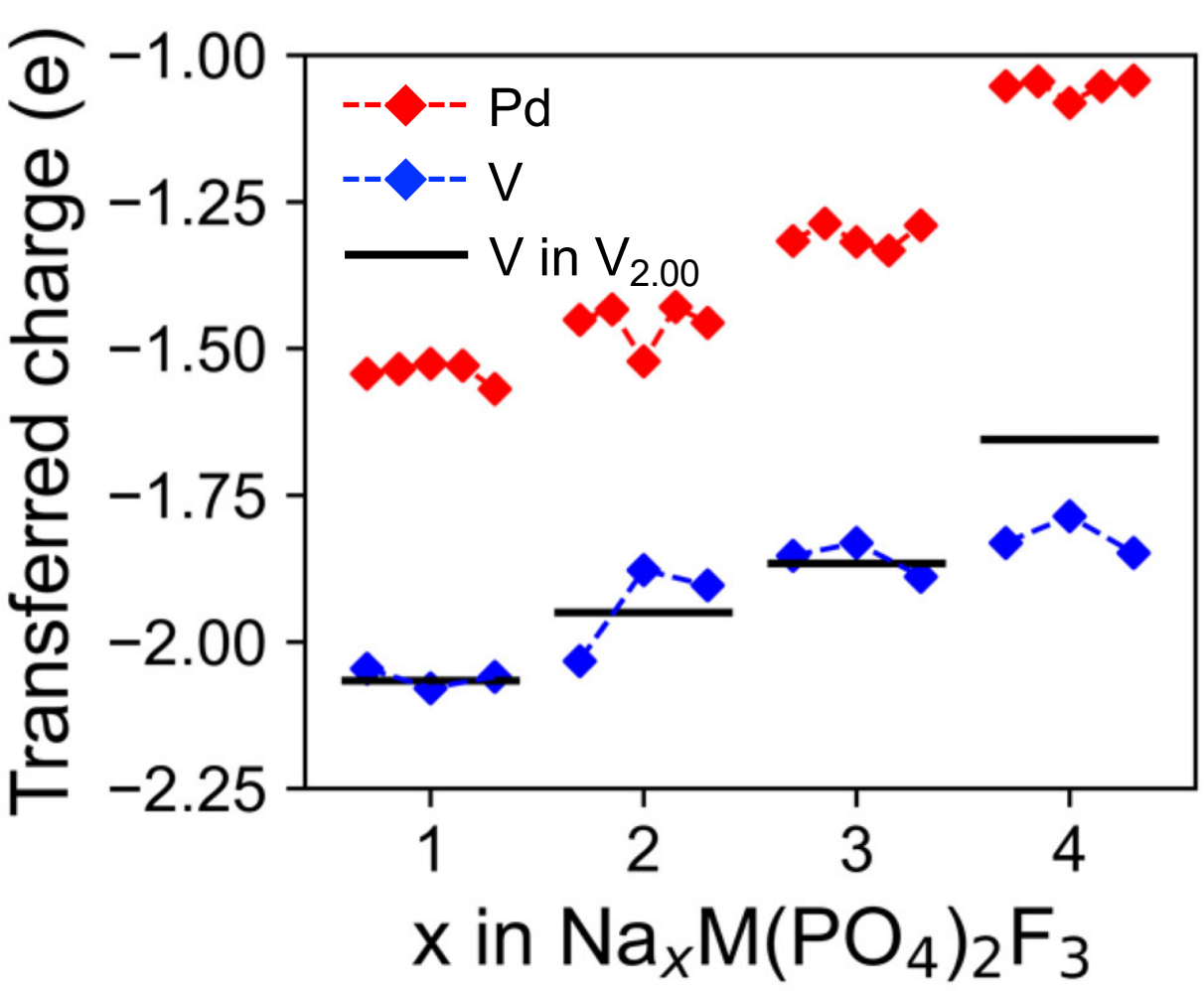

**Supplementary Fig. 11 | Atomic Bader charge analysis for transferred charge to $Pd_{1.25}V_{0.75}$ in NMPF.**

## Computational details

Using the Vienna Ab initio Simulation Package (VASP)[30] with the projector-augmented wave (PAW) scheme and spin-polarization, calculations were conducted following the Perdew–Burke–Ernzerhof (PBE) formulation[31]. A plane-wave expansion of wave functions was set with a cutoff energy of 500 eV, and a Monkhorst–Pack method[32] employing a 1×1×1 k-point grid was utilized. Convergence criteria for electronic and ionic steps in DFT calculations during observations in Bayesian optimization were $1.0\times10^{-4}$ eV and $1.0\times10^{-03}$ eV, respectively. For the validation of the NMPF with discovered element combinations, the convergence criteria were set at $1.0\times10^{-6}$ eV and -0.02 eV/Å. Hubbard U terms[33] were applied to V, Cr, Mn, Fe, Co, Ni, Cu and Zn for 3.1, 3.5, 4.0, 4, 3.3, 6.4, 4.0 and 7.5, respectively[34, 35]. The crystal structures were drawn with VESTA[36] software. In the Bayesian optimization algorithm, Gaussian process[24] which is a probabilistic machine learning model was used as the surrogate model with Matérn[11] kernel and an upper confidence bound (UCB) algorithm[37] is adopted for the acquisition function, which determines the maximum value of upper bound on uncertainty from Gaussian process regression.